\pdfoutput=1
\documentclass[twocolumn]{article}   
\usepackage{graphics} 
\usepackage{epsfig} 
\usepackage{mathptmx} 
\usepackage{times} 
\usepackage{amsmath} 
\usepackage{amsthm}
\theoremstyle{plain}
\newtheorem*{assumption*}{\assumptionnumber}
\providecommand{\assumptionnumber}{}


\usepackage{amssymb}  

\usepackage{epstopdf}
\usepackage{color}
\usepackage{soul}
\usepackage{comment}
\usepackage{ifthen}

\definecolor{darkred}{RGB}{144,9,9}
\newboolean{corron}
\setboolean{corron}{false} 
\newcommand\redst[1]{\ifcorron \textcolor{darkred}{\st{#1}} \fi}

\definecolor{darkblue}{RGB}{23,16,226}
\ifcorron
  \newcommand\blue{\textcolor{darkblue}}
\else
  \newcommand\blue{\textcolor{black}}
\fi

\date{}

\title{\LARGE \bf
Online wheel speed filtering for \redst{advanced bicycle applications}\blue{periodic disturbance reduction: a strategy for an advanced bicycle application}
}

\author{Gianmarco Rallo, Simone Formentin, Sergio M. Savaresi
\thanks{The authors are with Dipartimento di Elettronica, Informazione e Bioingegneria, Politecnico di Milano, via G. Ponzio 34/5, 20133 Milano (Italy). Email to: gianmarco.rallo@polimi.it.}%
}

\begin{document}
\maketitle
\thispagestyle{empty}
\pagestyle{empty}
\begin{abstract}
Due to geometrical errors and possible misalignment of the sensors, wheel speed measurements provided by incremental encoders in road vehicles are usually affected by significant periodic noises. This paper presents an online wheel speed filtering procedure, based on a model of the sensor, aimed at processing the speed measurement to make it suitable for advanced vehicle dynamics applications. \redst{, with a special focus}\blue{In particular, differently from low-pass and notch filtering, this strategy is reliable for the cycling cadence estimation from the wheel speed} on bicycles. Experimental data are used to show the effectiveness of the proposed approach.
\end{abstract}
\section{Introduction} \label{sec: Introduction}
In modern road vehicles, one of the most widely used sensor for measuring the speed of the wheels is the discrete incremental encoder. Among all the available technologies, magnetic encoders are generally preferred due to their compactness and robustness \cite{treutler2001magnetic}.

The working principle of magnetic encoders is simple: some permanent magnets assembled on a rigid disk (the so-called \emph{magnetic wheel}) rotate together with the wheel, while a Hall sensor (fixed on the vehicle chassis) transforms the magnetic field perturbed by the magnets into a voltage signal correlated with the original speed. An estimation problem then arises: the reconstruction of the wheel speed from the sequences of pulses given by the Hall sensor.
 
Such a problem has been extensively studied in the signal processing literature (see, \textit{e.g.}, \cite{kavanagh2001performance}, \cite{lemkin1995velocity} and \cite{su2005simple}), where it has been shown that the performance of each algorithm strongly depends on the considered ranges of speed and acceleration, see \cite{bascetta2009velocity} and \cite{phillips2003velocity}. However, almost all the methods can be seen as refinements of two basic techniques, both relying on a known (usually equispaced) geometrical allocation of the magnets along the circumference of the magnetic wheel: the \textit{pulses per period} algorithm, where the number of pulses within a fixed time interval defines the corresponding covered angle, and the \textit{fixed position} algorithm that computes the time interval between two consecutive pulses. With modern electronics, the time span defined by pulse detection events can be computed very accurately, thus the fixed position strategy is usually preferred to have a higher-resolution speed measurement.

Nevertheless, the experimental analysis in \cite{corno2010experimental} reveals the presence of a periodic noise affecting the reconstructed wheel speed when the fixed position algorithm is employed. In particular, it is shown that the most significant harmonics of the noise arise at multiples of the fundamental rotational frequency of the wheel. The analysis of \cite{panzani2012periodic} shows that such a noise could be attributed to the fact that the center of the encoder and the wheel rotation axis do not coincide.

The magnitude of such disturbances is usually large and has to be reduced, especially if the wheel speed measurement is employed for safety-critical applications like braking or traction control. 

To this aim, the authors of \cite{panzani2012periodic} show that adaptive notch filtering of the periodic disturbances should be preferred to simpler low-pass filtering, as the latter might introduce critical phase delay and remove some of the relevant features of the signal. However, ideal notch filtering is not feasible and, with real filters, some important \redst{information}\blue{features} in the signal could still be canceled together with the noise. 

\blue{This informations loss is critical in the application considered in this paper: in order to retrieve the cycling cadence from the wheel speed measurement it is necessary to preserve the oscillation which reveals the pedalling rate (see \cite{cadence2015}). Both low-pass and notch filtering are then not suitable in this case.}

Another \blue{filtering} approach is\redst{then} proposed in \cite{persson2002event,gustafsson2007tire}, under the assumption that the periodic disturbance is due to the non-ideal equispacing of the magnets on the magnetic wheel. The main idea of \cite{persson2002event} is to estimate these geometrical errors (first characterized in \cite{poisson1994measuring}) and compensate them in the fixed position strategy.

In this paper, the approach in \cite{persson2002event} is shown to be equivalent to a constrained least squares problem and an online implementation of the algorithm is proposed\blue{.}\redst{, with a focus on advanced bicycle applications} In particular, \redst{it will be shown that the proposed}\blue{this} strategy permits to \redst{preserve}\blue{selectively remove the periodic noise without compromising} the information about the cadence \redst{(\textit{i.e.} the pedal frequency)}contained in the speed signal and \redst{useful}\blue{needed} for cadence estimation \blue{see \cite{cadence2015})}. 

In terms of applicability, the method is based on \blue{recursive} constrained least squares and is therefore simple and computationally efficient. In addition, it is worth noting that the proposed strategy is suitable for any type of encoders, provided that the fixed position algorithm is used.

The paper is structured as follows. The algorithm and the implementation notes are provided in Section \ref{sec: Analysis of the method}. The experimental set-up and the filtering results are presented in Section \ref{sec: Experimental set-up and results}. \blue{A comparison with the notch filtering can be found in Section} \ref{sec: Comparison filtering}. The paper is ended by some concluding remarks.

\section{The filtering method} \label{sec: Analysis of the method}
Consider a magnetic encoder with $L$ magnets in a North/South ($N$/$S$) layout. It is here assumed that both rising and falling edges of the Hall sensor digital output \redst{cause}\blue{count as} a pulse, which is \redst{then}read by exploiting a $0\,V$ threshold \blue{on the original analog output}: if the voltage is positive (\textit{i.e.}, if a $N$-oriented magnet is exposed to the sensor), the digital output is $1$; as soon as a $S$-oriented magnet faces the sensor, the digital output becomes $0$. 

Without loss of generality, the delays of the Hall sensors as well as the quantization error in the measurement of the time between two pulses are here supposed to be negligible, and the rotational speed is assumed to be $\omega(t) > 0,\,\, \forall t$ (changes of the rotation direction are not contemplated). Notice that, under the above assumptions, $L$ pulses occur per each revolution.

The basic version of the fixed position algorithm relies on two additional assumptions: (i) the magnets are equispaced and (ii) they are identical. Therefore, in the time interval defined by two consecutive pulses, the magnetic wheel (jointly with the vehicle's wheel) covers a constant angular distance equal to
\begin{equation} \label{eq: alpha nom}
\alpha_{\mathrm{nom}} = \frac{2\pi}{L}.
\end{equation}
The speed measurement is updated when a new pulse is observed: this approach is usually referred to as \textit{event based sampling}. If $t_{i,k}$ indicates the instant of occurrence of the $i^{\mathrm{th}}$-pulse out of $L$ total events expected for the $k^{\mathrm{th}}$ revolution and $\Delta t_{i,k} = t_{i,k}-t_{i-1,k}$ is the elapsed time from the previous event, the computation of the value of the speed reads
\begin{equation} \label{eq: omega basic}
\omega^{\mathrm{b}}(t_{i,k}) = \omega^{\mathrm{b}}_{i,k} := \frac{\alpha_{\mathrm{nom}}}{\Delta t_{i,k}}.
\end{equation}
Such algorithm holds for all types of encoder, provided that any rotating part (\textit{e.g.} the magnetic wheel) can be modelled as a disk partitioned by \textit{marks} (\textit{e.g.} magnets) into $L$ equal circular sector of width $\alpha_{\mathrm{nom}}$.

According to \cite{persson2002event}, when the magnets are not exactly equispaced, the fixed point algorithm yields a biased result since the width of the circular sectors is no longer constant and equal to $\alpha_{\mathrm{nom}}$. Similar errors could be induced by different reasons, \textit{e.g.} if the axis of the Hall sensor is not effectively perpendicular to the plane of the magnetic wheel (which actually can be not perfectly planar) or if the encoder center and the rotation axis of the vehicle's wheel do not coincide \cite{panzani2012periodic}.

\begin{figure}
\centering
\includegraphics[width = 0.6\columnwidth]{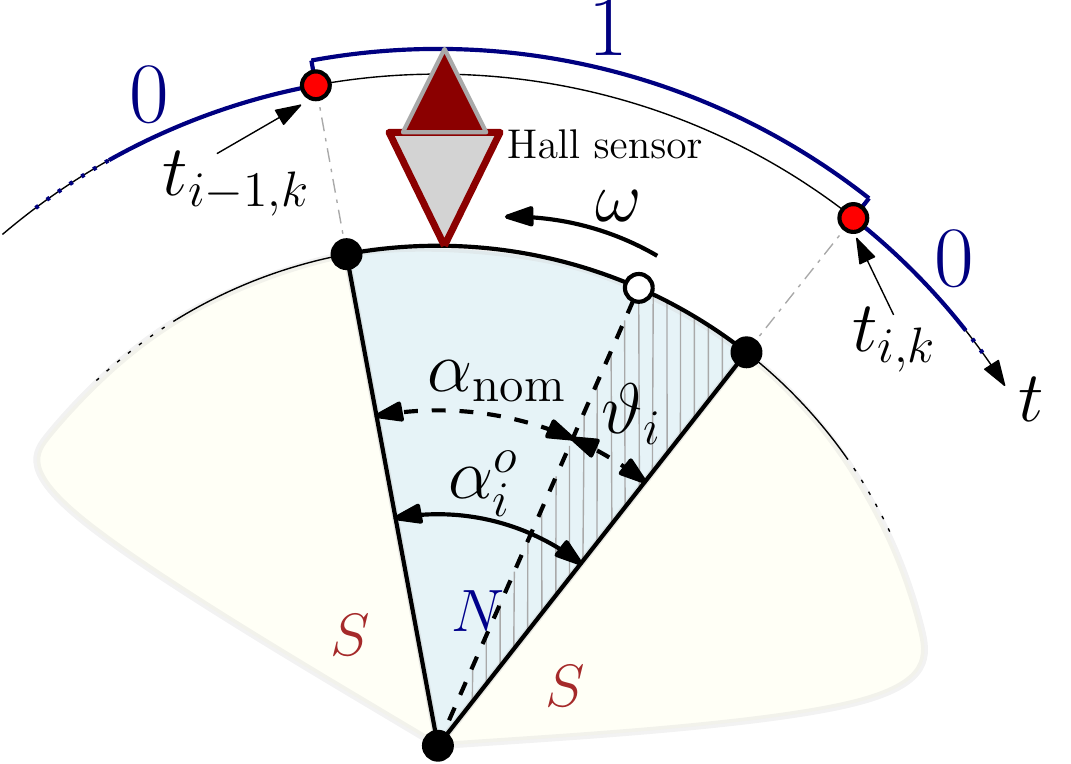}
\caption{A scheme of the magnetic encoder, assuming that the $i^{\mathrm{th}}$ sector refers to a $N$-oriented magnet.}
\label{fig:single_sector}
\end{figure}

In Figure \ref{fig:single_sector}, a graphical representation of a realistic setting is shown: the $i^{\mathrm{th}}$ sector (\textit{i.e.} the portion spanned in the time interval $\left[ t_{i-1,k},t_{i,k} \right],\,\,\forall k$) is facing the Hall sensor. A measurement error $e_{i,k}$ is originated from the depicted equivalent geometrical error $\vartheta_{i}$ as
\begin{equation}\label{eq: measurement error}
\omega^{\mathrm{b}}_{i,k} = \frac{\alpha_{\mathrm{nom}}}{\Delta t_{i,k}} = \frac{\alphaì^{o}_{i}-\vartheta_{i}}{\Delta t_{i,k}} = \omega^{o}_{i,k} +  e_{i,k},
\end{equation} 
where $\omega_{i,k}^{o}=\frac{\alpha^{o}_{i}}{\Delta t_{i,k}}$ indicates the true rotational speed.

From \eqref{eq: measurement error}, it can be observed that the measurement of a constant rotational speed (in case of $\omega^{o}(t) = \bar{\omega}\, \forall
t$ and supposing $\vartheta_{i}\neq 0\,\,\forall i \in \lbrace 1,\dots,L \rbrace$) would suffer from periodic disturbances at $\lbrace i  \bar{\omega} \rbrace_{i=1}^{L}$, \textit{i.e.} the multiples of the \textit{fundamental} angular frequency $\bar{\omega}$. 

In \cite{persson2002event}, it is observed that, by considering the true covered angular distance $\lbrace \alpha_{i}^{o} \rbrace_{i=1}^{L}$ sector-by-sector instead of the fixed $\alpha_{\mathrm{nom}}$ and compensating an estimates of the geometrical errors $\vartheta_{i}$, the quality of the speed measurement can be further improved. In what follows, we elaborate upon this idea, by reformulating the estimation/compensation issue as a constrained Least Squares (cLS) problem and proposing an efficient online implementation.

First of all, assume that the unknown true speed $\omega_{i,k}^{o}$ is well approximated by the mean revolution speed $\omega^{\mathrm{rev}}_{i,k}$, which is instead known and available once per revolution without geometrical errors. In particular,   
\begin{equation} \label{eq: main hypothesis}
\omega^{o}_{i,k} = \frac{\alpha^{o}_{i}}{\Delta t_{i,k}} = \frac{\alpha_{\mathrm{nom}}+\vartheta_{i}}{\Delta t_{i,k}} \approx \omega^{\mathrm{rev}}_{i,k}=\frac{2\pi}{t_{i,k}-t_{i,k-1}}=\frac{2\pi}{\Delta t^{\mathrm{rev}}_{i,k}},
\end{equation}
where $\Delta t_{i,k}^{\mathrm{rev}}$ indicates the revolution time evaluated referring to the $i^{\mathrm{th}}$ event. This is achieved exactly for constant speed.

It is then possible to accomplish the estimate of the geometrical errors in several ways. Here, the implementation is based on $L$ linear regression models (updated once per revolution), which are derived for each sector with the scope of stating the equivalent $\alpha^{o}_{i}$ estimation problem (easily derived from \eqref{eq: main hypothesis}):
\begin{equation} \label{eq: regression ith sector}
y_{i}(k) = \varphi^{T}_{i}(k) \alpha^{o}_{i},\, i =1,\dots,L,
\end{equation}
where $\varphi(k) = 1\,\forall k$ and $y_{i}(k)=\frac{2\pi \Delta t_{i,k}}{\Delta t^{\mathrm{rev}}_{i,k}}$. 
Furthermore, sectors angular widths $\alpha^{o}_{i}$ (which are positive, according to models \eqref{eq: regression ith sector}) satisfy the obvious geometrical constraint
\begin{equation} \label{eq: geometrical constraint}
\sum\limits_{i=1}^{L} \alpha^{o}_{i} = 2\pi.
\end{equation}

Hence, when a set of $N$ revolutions is available, the estimates of the errors are given as the solution of the simple cLS problem
\begin{equation}\label{eq: batch problem formulation}
\begin{split}
\min\limits_{\underline{\alpha}^{o} \in \mathbb{R}^{L\times 1}}& \sum\limits_{k = 1}^{N} \left( \Phi^{T}(k) \underline{\alpha}^{o}-\underline{Y}(k) \right)^{2}\\
\text{s.t.} \quad &A\underline{\alpha}^{o} = b
\end{split}
\end{equation}
where $\underline{\alpha}^{o} = \left[ \alpha^{o}_{1},\dots,\alpha^{o}_{L} \right]^{T}$ is the vector of unknowns, $\underline{Y}(k) = \left[ y_{1}(k),\dots,y_{L}(k) \right]^{T} \in \mathbb{R}^{L\times 1}$ is the vector of observations at $k^{\mathrm{th}}$ ripple, $\Phi^{T}(k) = I_{L}\, \forall k$ is the matrix form of the constant regressor (and $I_{L}$ indicates the identity matrix of size $L$), while $A = [1,\dots,1]\in \mathbb{R}^{1 \times L}$ and $b = 2\pi$ express the geometrical constraint in \eqref{eq: geometrical constraint}.

The closed-form solution of a cLS problem (when the constraint is linear) is derived in \cite{kay1993fundamentals} and can be seen as a suitable redistribution of the constraint violation errors of the unconstrained solutions over the dimensions of the unknown. In this specific case, the batch cLS estimate reads
\begin{equation}\label{eq: constrained batch}
\hat{\underline{\alpha}}^{o}_{N} = \hat{\underline{\alpha}}^{\mathrm{unc}}_{N} -A^{T} \cdot \left[ \frac{1}{L} \left( A\hat{\underline{\alpha}}^{\mathrm{unc}}_{N}-b \right) \right],
\end{equation}
where 
\[\hat{\underline{\alpha}}^{\mathrm{unc}}_{N} = \frac{1}{N} \left[ \sum\limits_{k=1}^{N} \Phi(k)\Phi^{T}(k) \right]^{-1} \sum \limits_{k=1}^{N} \underline{Y}(k) = \frac{1}{N} \sum \limits_{k=1}^{N} \underline{Y}(k)\]
indicates the unconstrained batch LS solution with the provided $N$ data. Notice that the violation error turns out to be divided into $L$ equal parts.

We propose next a recursive formulation of \eqref{eq: constrained batch} by following the same rationale:
\begin{subequations} \label{eq: recursive estimate}
\begin{align}
&n(k) = \mu n(k-1) + 1 \label{eq: recursive estimate sub 1} \\
&\hat{\underline{\alpha}}^{\mathrm{unc}}(k) = \hat{\underline{\alpha}}^{\mathrm{unc}}(k-1) + \frac{1}{n(k)} \left[\underline{Y}(k)-\hat{\underline{\alpha}}^{\mathrm{unc}}(k-1) \right] \label{eq: recursive estimate sub 2} \\
&\hat{\underline{\alpha}}^{o}(k) = \hat{\underline{\alpha}}^{\mathrm{unc}}(k) + \frac{1}{L} \left[ A\hat{\underline{\alpha}}^{\mathrm{unc}}(k)-b \right].\label{eq: recursive estimate sub 3}
\end{align}
\end{subequations}
where $\mu <1$ denotes the forgetting factor \cite{young2012recursive}.
Notice that \eqref{eq: recursive estimate} is composed by two parts: the recursive estimation in \eqref{eq: recursive estimate sub 1} and \eqref{eq: recursive estimate sub 2} of the unconstrained solution and the adjustment \eqref{eq: recursive estimate sub 3} to ensure the fulfillment of the constraint. The inverse of the variable $n(k)$ weights the last datum; in particular
\begin{equation}\label{eq: Nw definition}
N_{w} := \lim\limits_{k\to\infty}n(k) = \frac{1}{1-\mu},
\end{equation}
represents the length of the window including significant data for the estimation \cite{young2012recursive}. This also means that, after $N_{w}$ revolutions, a misleading evaluation of the errors can be suitably corrected.

The final model-based filtered speed $\omega^{\mathrm{f}}$ is then a refinement of the computation of the speed for each sector, with the correct $\hat{\underline{\alpha}}^{o}(k)$, namely
\begin{equation}\label{eq: final formulation}
\omega^{\mathrm{f}}_{i,k}=\frac{\hat{\alpha}^{o}_{i}(k)}{\Delta t_{i,k}} \quad i =1,\dots,L.
\end{equation}
\section{Experimental results} \label{sec: Experimental set-up and results}
\begin{figure}
\centering
\includegraphics[width = 0.8\columnwidth]{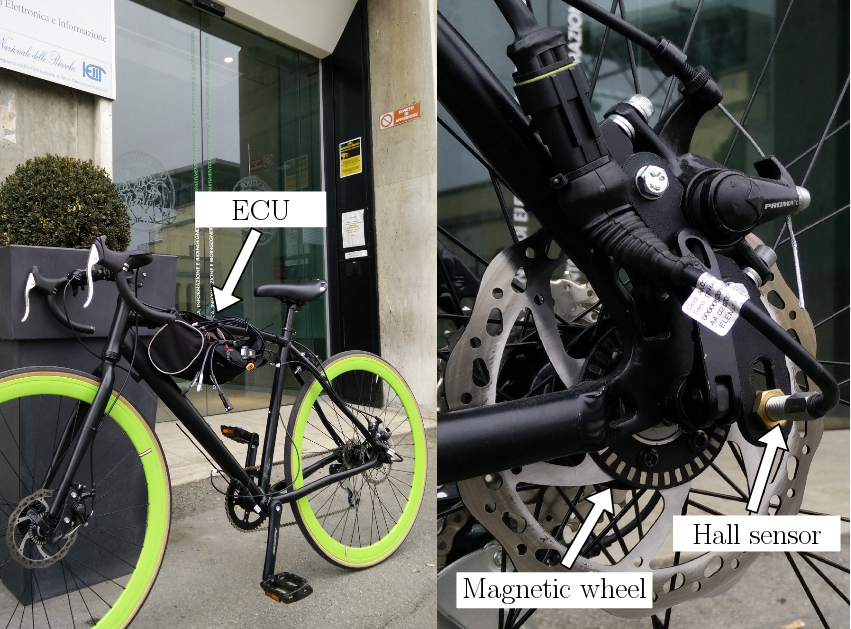}
\caption{The employed experimental set-up: a racing bike equipped with a magnetic encoder and an ECU.}
\label{fig:bike}
\end{figure}

In order to experimentally validate the performance of the proposed filtering strategy, the multi-geared bicycle with racing frame (wheel radius $R = 0.334\,m$) of Figure \ref{fig:bike} is employed. On such a vehicle, it will be possible to show the advantages of the proposed approach for advanced bicycle applications.

The vehicle is equipped with a rotational speed measurement apparatus consisting of: (i) a magnetic encoder (Hall sensor plus magnetic wheel with $L=36$ permanent magnets in likewise slots with a \textit{N/S layout}, thus $\alpha_{\mathrm{nom}}=10^{\circ}$); (ii) a compact ECU (\textit{Electronic Control Unit}) devoted to the online speed computation and filtering, along with its power source (a $12V$ battery pack); (iii) a CAN-Bus that defines the communication interface among the listed entities.
A CAN data-logger is finally used for collecting and saving the experimental results.

\begin{figure}
\centering
\includegraphics[width = 0.8\columnwidth]{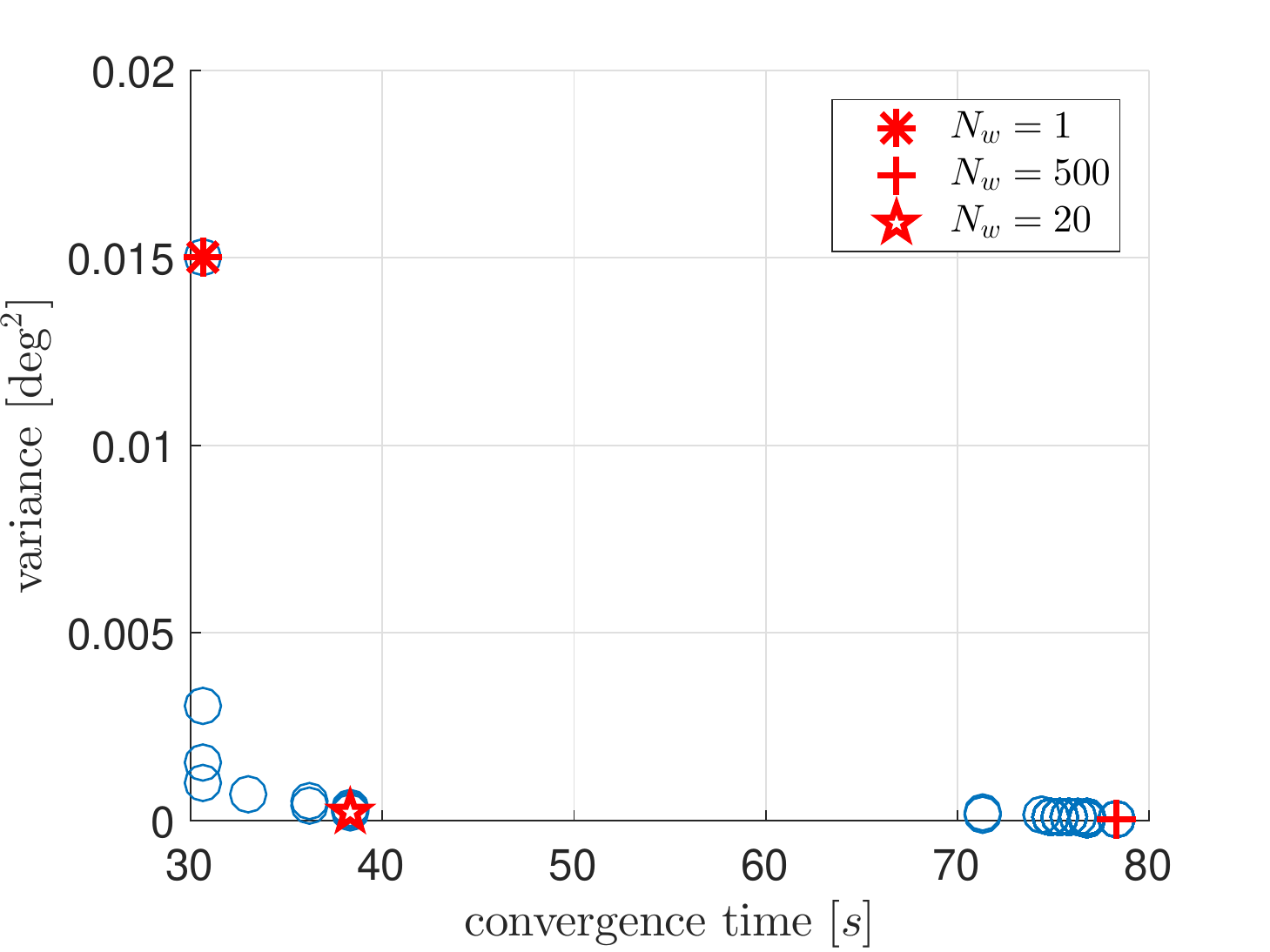}
\caption{Experimental trade-off diagram for the choice of $N_w$: a balance between variance and convergence time.}
\label{fig: Nw tradeoff}
\end{figure}

The choice of $N_{w}$ is made \emph{a-priori} as a good trade-off between accuracy and convergence time (both increasing with $N_{w}$). The former is quantified using the mean variance of the $L$ adaptive estimates while the convergence time is set as the time needed to obtain an absolute error (with respect to the corresponding offline batch solutions) of $5\%$. The trade-off diagram depicted in Figure \ref{fig: Nw tradeoff} shows that a suitable choice is $N_{w}=20$ (revolutions). 

For bicycle applications, the convergence time is particularly advantageous. In fact, the encoder is not able to provide the rotating direction, thus possible small inverse rotations of the wheel (which can occur when the cyclist stops) might compromise the quality of the estimation if not recovered in small time. For the same reason, the estimate of the geometrical errors is enabled only when the speed is greater than $5\,\frac{km}{h}$, while it is reset if the speed is lower.

\begin{figure}
\centering
\includegraphics[width = 0.8\columnwidth]{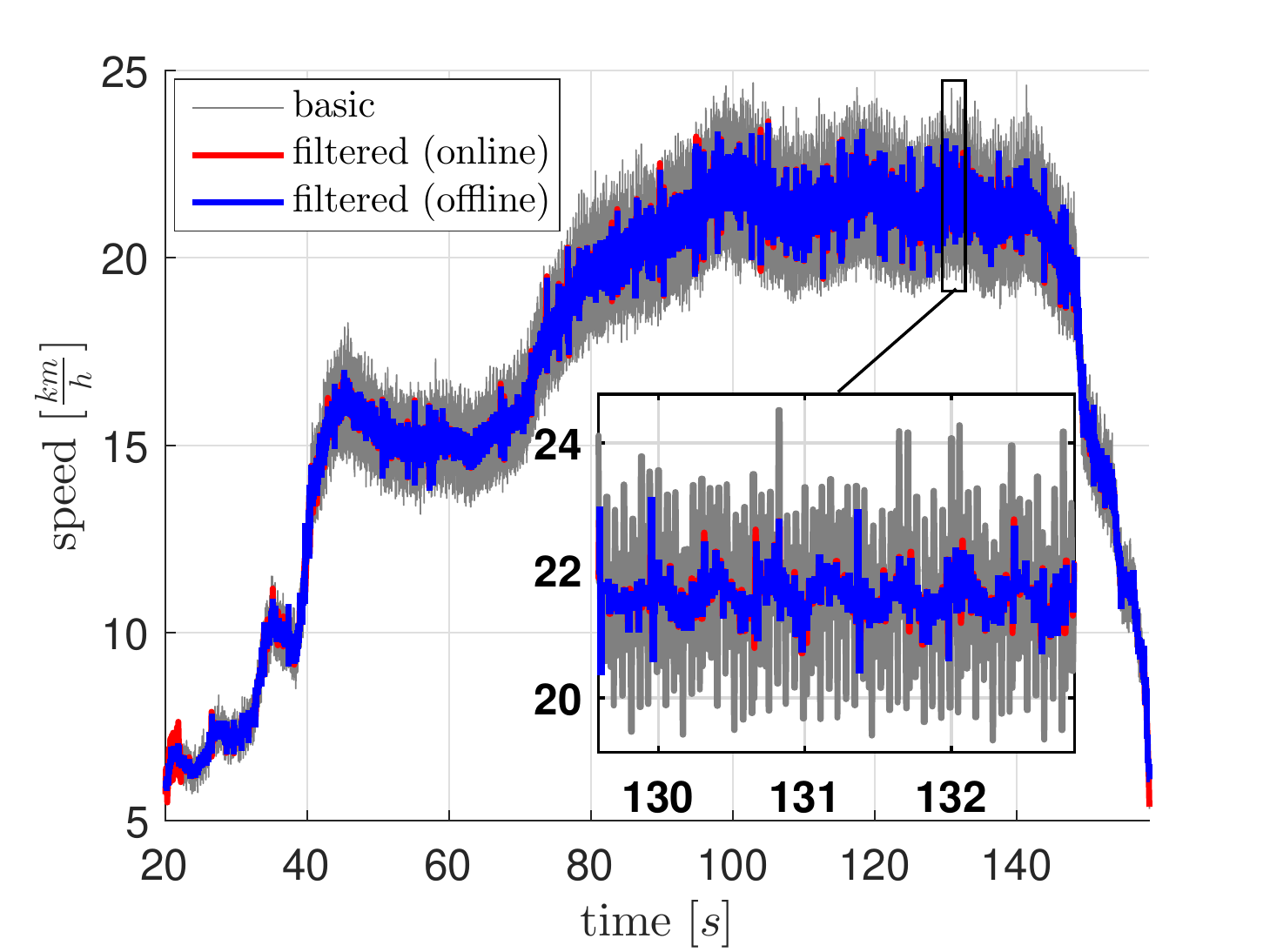}
\caption{Wheel speed data: raw, offline filtered and online filtered.}
\label{fig: time domain}
\end{figure}

The results achieved in the time domain by the proposed speed filtering strategy, along with the measurement via the basic fixed position algorithm, are presented in Figure \ref{fig: time domain} (where the linear speed, expressed in $\frac{km}{h}$, is directly inferred from $\omega^{\mathrm{f}}$ and the wheel radius $R$) and refer to a test performed on a straight road (dry asphalt) by a non-professional cyclist. Pedals were \textit{engaged} for the whole test and there was no gear shifting. Notice that the online filtered signal is very similar to the offline estimation of \cite{persson2002event} (using the entire data-set). For completeness, the results are such that $\left| \hat{\vartheta}_{i} \right| = \left| \alpha_{i}^{o}-\alpha_{\mathrm{nom}} \right| \in \left[0.05,0.98 \right]^{\circ}\,\forall i=1,\dots,L$  while the mean absolute value is equal to $0.44^{\circ}$.

\begin{figure}
\centering
\includegraphics[width = 0.8\columnwidth]{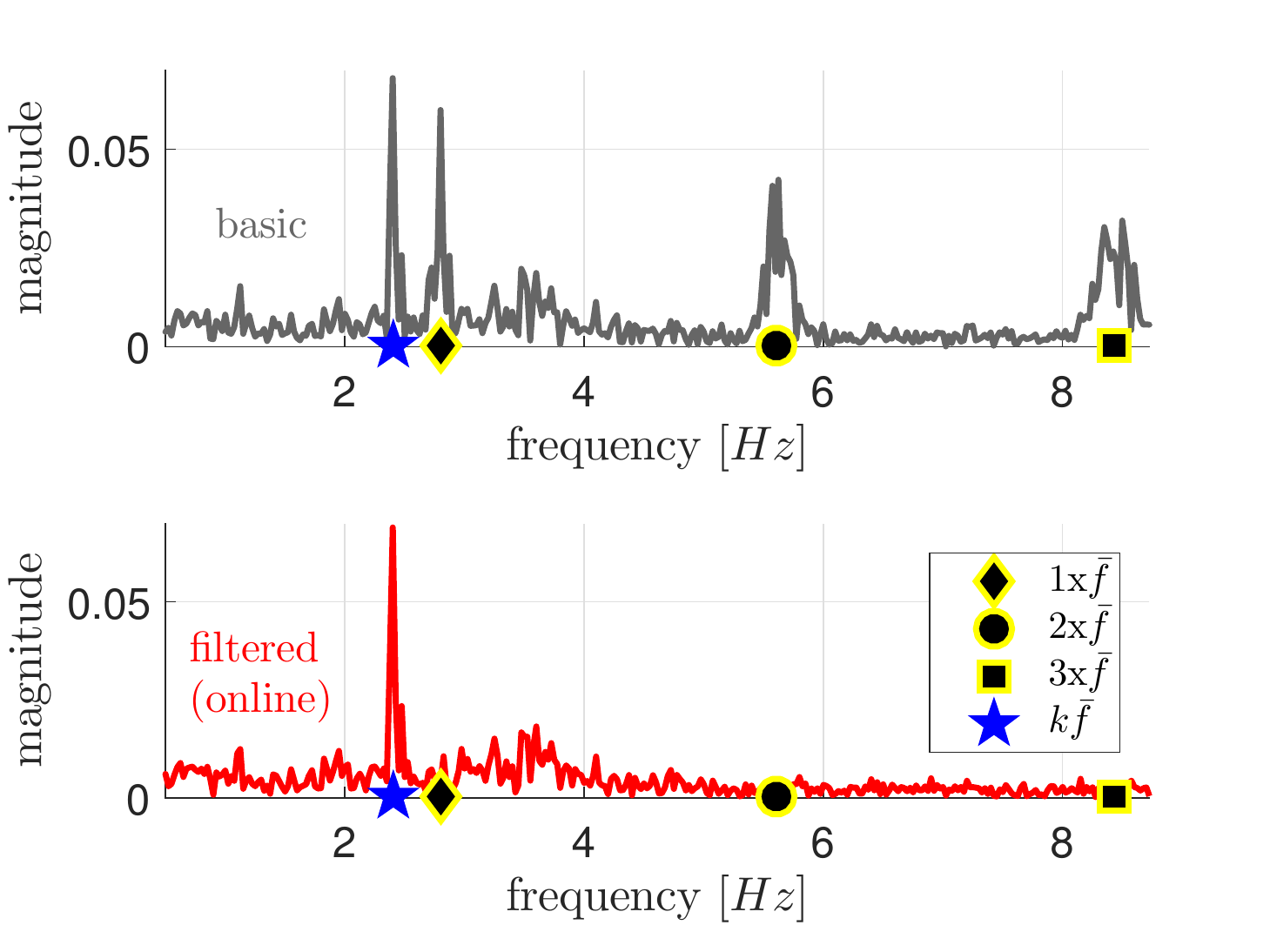}
\caption{Frequency-domain wheel speed data: with (bottom plot) and without filtering (top plot).}
\label{fig: freq domain}
\end{figure}

A clarifying picture of the advantages given by the method is shown in the frequency domain analysis of Figure \ref{fig: freq domain}. To obtain the figure, the data within the time interval $\left[100,140 \right]\,s$ (characterized by a quasi-constant speed) are employed. Notice that the \blue{mean} speed is approximately $21.4\,\frac{km}{h}$ within this window, \redst{that means}\blue{corresponding to} $\bar{\omega} = 17.64\,\frac{rad}{s}$ \redst{, or a frequency peak at}\blue{or} $\text{\blue{$\bar{f}=$}}2.8\,Hz$. As expected, the Fourier transform of the basic measurement reveals not only the presence of \redst{a peak}\blue{disturbance} at \redst{$2.8\,Hz$}\blue{$\bar{f}$ (indicated by the diamond mark)}, but also\redst{some disturbance} components at multiples of the fundamental \blue{rotational} frequency \blue{(circle and square marks display the position of the second and the third harmonic, respectively)}. These additional (undesired) terms are successfully removed through the proposed online model-based filtering (see the bottom plot of Figure \ref{fig: freq domain}). 

The remaining peak in the \blue{filtered} speed spectrum \redst{shows the cadence of the rider}\blue{(marked by the star) derives from the nature of the pedaling: most cyclists are not able to provide a constant torque throughout the pedal crank revolution but two peaks occur alternatively with two dead points based on the pedal position (see \cite{bertucci2012analysis}). Hence the torque oscillates with a frequency equal to twice the pedaling rate and it is almost rigidly transimtted to the wheel (thus to the speed measurement) through the chain. This assumption can be easily verified knowing the transmission gain: the position of this peak coincides with $k\bar{f}=2.41\,Hz$, where $k=0.8607$ corresponds to twice the gain of the adopted gear ratio.} \redst{c, }\blue{A}s discussed in \cite{cadence2015}, \redst{that}\blue{the cadence} can therefore be estimated using the wheel sensor without adding an additional encoder on the pedal. Notice that such a peak could be accidentally removed in some operating conditions if different methods (like the notch filtering of \cite{panzani2012periodic}) are used\blue{, as shown in the next section}. 

\begin{figure}
\centering
\includegraphics[width = 0.8\columnwidth]{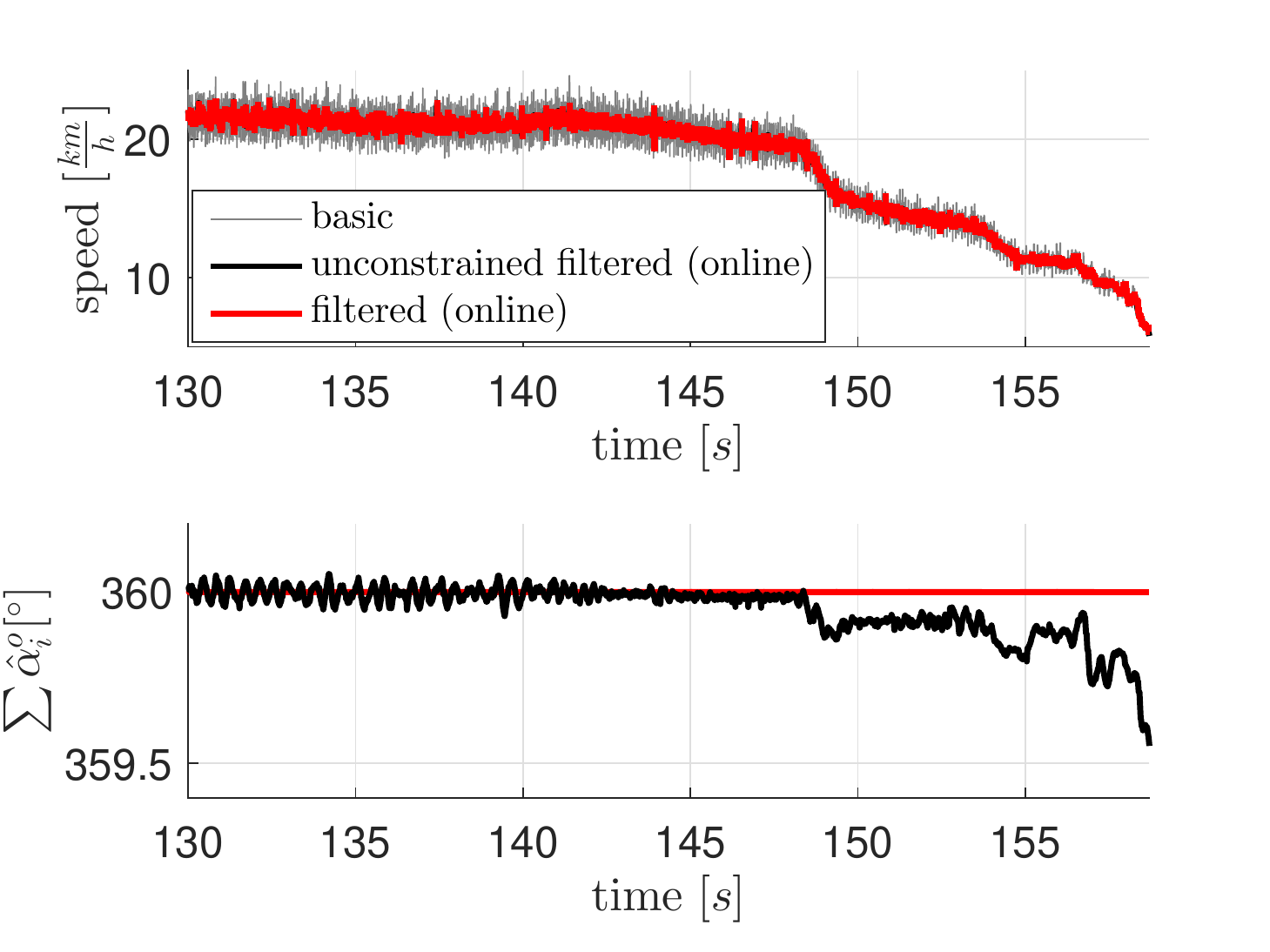}
\caption{During braking, the true speed is no longer well approximated by the mean revolution speed. If the geometrical constraint is not accounted for, the errors estimate becomes unreliable.}
\label{fig: unconstrained}
\end{figure}

For completeness, notice that the assumption that the true speed is well approximated by the mean revolution speed is not verified when the rotational speed varies significantly within a single revolution, \textit{e.g.} during sudden braking or accelerating phases. Nevertheless, we observed that the adaptive nature of the algorithm helps to recover the correct estimates in reasonable time and the additional information provided by the geometrical constraint \eqref{eq: geometrical constraint} strongly limits the error introduced by the violation of the hypothesis. In Figure \ref{fig: unconstrained}, some estimation results using an unconstrained estimate, as well as the corresponding constraint violation error during the braking phase of the considered test, are illustrated to further certify the above statement.

\section{\blue{Comparison with notch filtering}} \label{sec: Comparison filtering}
The notch filtering strategy proposed in \cite{panzani2012periodic} allows to efficiently remove the periodic disturbance in the wheel speed measurement and outperforms the simple low-pass filtering in terms of phase-delay of the resulting signal. However, the main drawback of this approach is that any (possibly useful) information located at the notched frequencies are filtered along with the periodic disturbance components.

This is the case of the considered application: in particular, the pedaling-related speed oscillation arises in the neighborhood of $\bar{f}$. This assumption holds given the commonly adopted gear ratios in bicycles, \textit{e.g.} for the multi-geared bike used for the experimental analysis $k\in\left[0.57,1.33\right]$.

\begin{figure}
\centering
\includegraphics[width = 0.8\columnwidth]{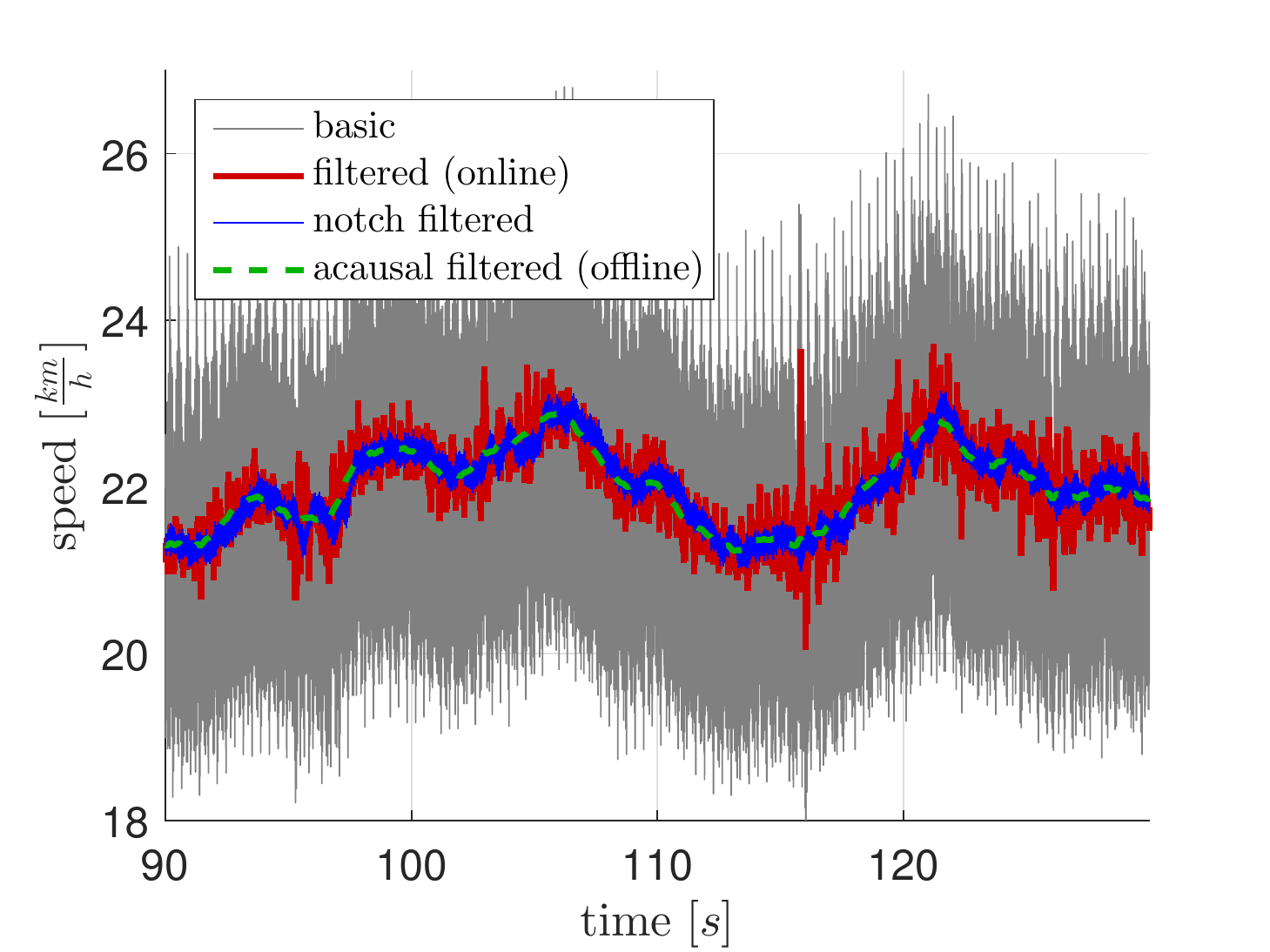}
\caption{Filtering strategies comparison: time domain results.}
\label{fig: filter comparison time}
\end{figure}

\begin{figure}
\centering
\includegraphics[width = 0.8\columnwidth]{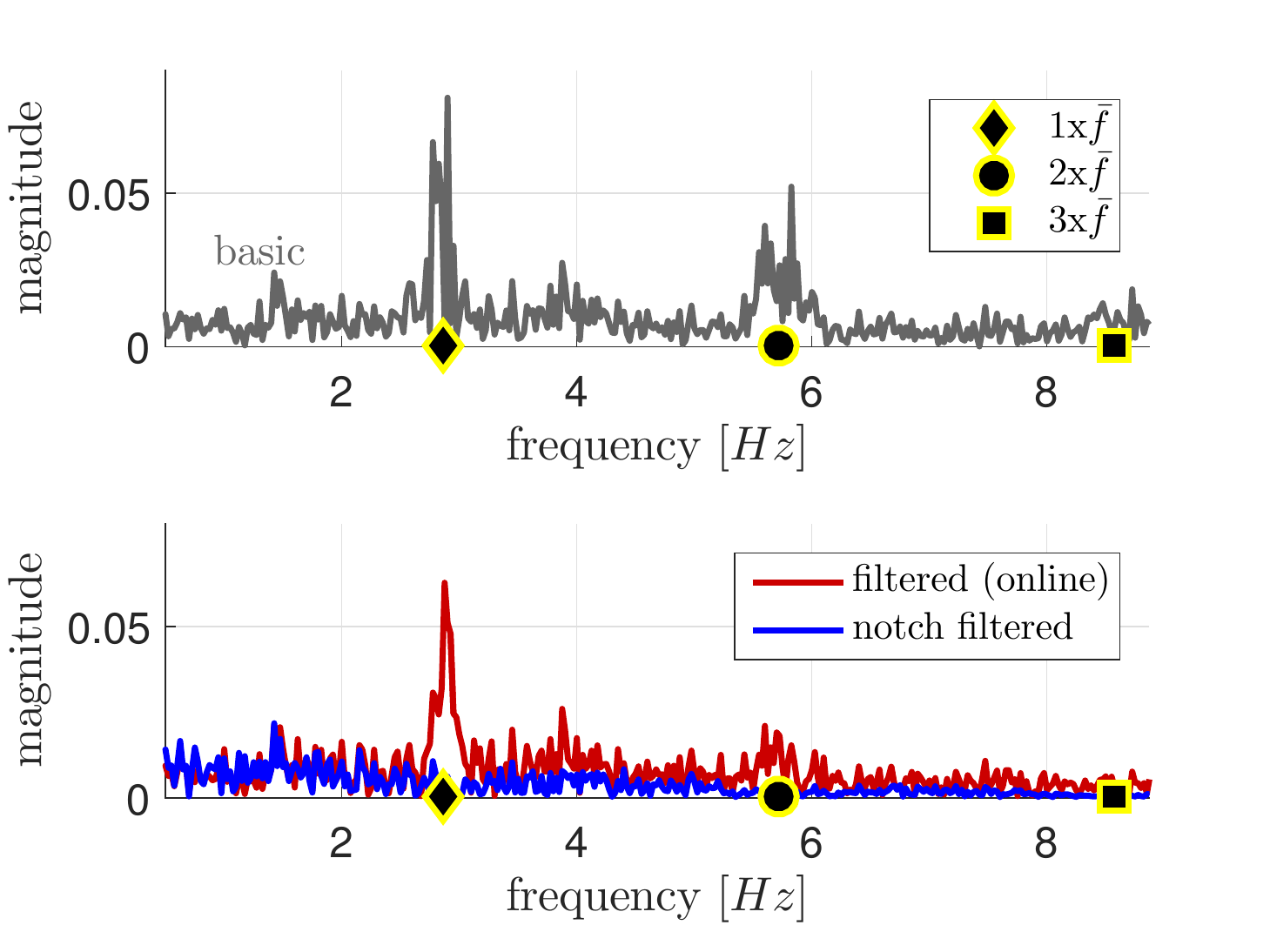}
\caption{Filtering strategies comparison: frequency domain results.}
\label{fig: filter comparison frequency}
\end{figure}

In order to prove the suitability of the proposed strategy, a test with $k=1$ (it is an available value in the gearbox) has been performed: in this case the frequency of the torque ripples coincides with $\bar{f}$.
Notice that the low-pass filtering approach is not suitable for this application and it has not been included in the analysis.
The results in the time domain are shown in Figure \ref{fig: filter comparison time}: the outputs provided by the considered strategies are compared with the acausal low-pass filtered version of the speed (computed offline). Performance provided by the notch filter are better considering the overall speed signal \emph{quality}, but the information about the cadence has been lost. On the contrary, the proposed model-based strategy preserves the pedaling oscillation while reducing the effect of the periodic disturbance. This achievement is more evident in the frequency domain as illustrated in Figure \ref{fig: filter comparison frequency}.

\section{Conclusions} \label{sec: Conclusions}
In this paper, an online filtering strategy for the rotational speed measured via incremental encoders is discussed. The analysis points out that the online estimate of the geometrical errors, which by assumption cause the observed periodic disturbance, can be accomplished through simple recursive cLS. Experimental results on a racing bike confirm the theoretical expectations and the effectiveness of the approach for \redst{vehicle dynamics applications}\blue{selectively filtering the periodic noise. Due to the model-based nature of this strategy, useful informations contained in the speed signal, such as the cadence-related oscillation, are instead preserved}.

\section*{Acknowledgements} \label{sec: ack}
Fabio Todeschini and Blubrake srl are gratefully acknowledged for their technical support on the experimental tests.

\bibliographystyle{plain}
\bibliography{biblio_persson}

\end{document}